\documentclass[oneside,reqno,11pt]{amsart}
\usepackage{amsmath}
\usepackage{amsfonts}
\usepackage{amssymb}
\usepackage{amsthm}
\usepackage{graphicx}
\usepackage{float}
\usepackage{mathtools}
\usepackage{caption}
\usepackage{braket}
\usepackage[english]{babel}
\usepackage{enumitem}
\usepackage{csquotes}
\usepackage{xparse}
\usepackage{hyperref}
\hypersetup{
	colorlinks=true,
	linkcolor=Red4, 
	citecolor=blue
}

\usepackage{fontawesome}

\NewDocumentCommand{\overarrow}{O{=} O{\downarrow} m}{%
	\overset{\makebox[0pt]{\begin{tabular}{@{}c@{}}#3\\[0pt]\ensuremath{#2}\end{tabular}}}{#1}
}

\usepackage[backend=bibtex,giveninits=true,citestyle=numeric,bibstyle=numeric,maxbibnames =100,maxcitenames=3,isbn=false]{biblatex}
\AtBeginBibliography{
	
}

\renewbibmacro{in:}{}
\DeclareFieldFormat{pages}{#1}

\newtheorem{theorem}{Theorem} 

\newtheorem{lemma}{Lemma}

\theoremstyle{definition}
\newtheorem*{example*}{Example}
\newtheorem*{remark*}{Remark}
\usepackage[foot]{amsaddr}
 
\makeatother

\usepackage{palatino}
\usepackage{lettrine}
\usepackage{erewhon}
\usepackage{GoudyIn}

\usepackage[x11names]{xcolor} 

\LettrineTextFont{\itshape}
\newcommand{\R}{{\mathbb R}}
\newcommand{\Ce}{{\mathbb C}}
\newcommand{\N}{{\mathbb N}}
\newcommand{\I}{{\mathrm i}}

\newcommand{\e}{\operatorname{e}}
\newcommand{\tr}{\operatorname{tr}}
\newcommand{\spann}{\operatorname{span}}
\newcommand{\minus}{\scalebox{0.5}[1.0]{$-$}}
\newcommand{\mminus}{\scalebox{0.75}[1.0]{$-$}}

\usepackage[margin=3.5cm,bottom=3.5cm,top=3.5cm]{geometry}
\usepackage{tikz}
\usetikzlibrary{decorations.pathreplacing}
\makeatletter
\def\uunderbrace#1{%
	\@ifnextchar_{\tikz@@uunderbrace{#1}}{\tikz@@uunderbrace{#1}_{}}}
\def\tikz@@uunderbrace#1_#2{%
	\tikz[baseline=(a.base)] {\node[inner sep=0.5] (a) {\(#1\)};
		\draw[line cap=round,decorate,decoration={brace,amplitude=4pt}]
		(a.south east) -- node[pos=0.5,below,inner sep=6pt] {\(\scriptstyle #2\)} (a.south west);}}

\makeatother

\renewenvironment{abstract}
{\small
	\begin{center}
	\end{center}
	\list{}{%
		\setlength{\leftmargin}{4.0mm}
		\setlength{\rightmargin}{\leftmargin}%
	}%
	\item\relax}
{\endlist}

\begin{document}

	\title{Orthogonal Projections on Hyperplanes Intertwined With Unitaries}
	
	\author{\vspace{-2mm} Wojciech S\l omczy\'{n}ski$\ $}
	\author{$\ $Anna Szczepanek}
	\address{Institute of Mathematics, Jagiellonian University, \L ojasiewicza 6, 30-348 Krak\'{o}w, Poland}
	\email{{\scriptsize  wojciech.slomczynski@im.uj.edu.pl,$\!\!$ anna.szczepanek@uj.edu.pl}}

 	\vspace*{-11mm}
	\maketitle
	
	\begin{abstract}

		\vspace{-8mm}
		
\textsc{Abstract.} 
Fix a point in a finite-dimensional complex   vector space and consider the sequence of iterates of this point under the composition of 
a~unitary   map with the orthogonal projection on the hyperplane orthogonal to the starting point.  
We prove that, generically, the  series of the squared norms of these iterates sums to the dimension of the underlying space. 
This leads us to construct a (device-dependent) dimension witness for quantum systems  which involves the probabilities of obtaining certain strings of outcomes in a sequential yes-no measurement. 
 The exact formula for this series in non-generic cases is provided as well as its analogue in the real~case.

 		\vspace{0.25mm}
		

 		\vspace{0.25mm}
		
		 
		\vspace{0.5mm}
		
	\end{abstract}

	\maketitle
	\section{Results}
\begin{figure}[H]
	 	\vspace{-6mm}
		\centering
		\includegraphics[scale=0.525]{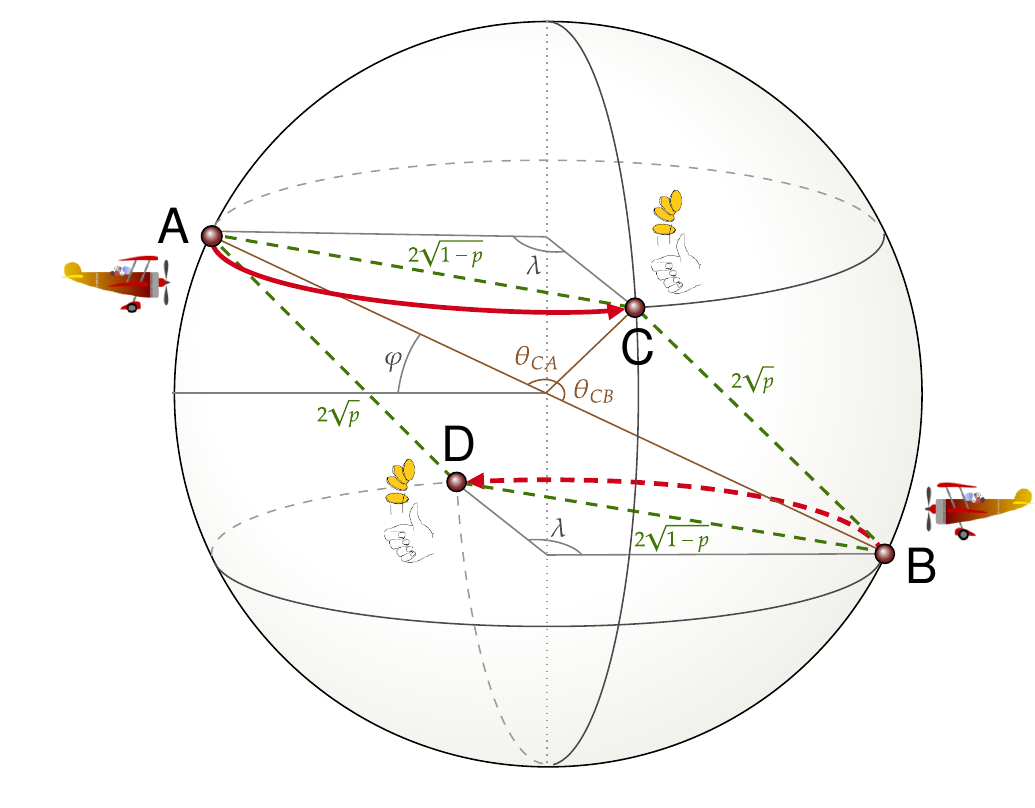}
		\vspace{-4mm}
		\captionsetup{width=0.775\linewidth}
		\caption{Red arrows represent Alice's eastward journey from $\mathsf{A}$ to $\mathsf{C}$ (or from $\mathsf{B}$ to~$\mathsf{D}$), where she has to decide  whether to go to $\mathsf{A}$ or $\mathsf{B}$. Green dashed segments show the  (straight-line) distances that separate Alice from her potential destinations.}\label{fig1}
	\end{figure}

	\thispagestyle{empty}
	\lettrine{I}{magine} 
	Alice and Bob live in two \textit{antipodal} cities, say Alaejos
	in Spain ($\mathsf{A}$ for Alice) and Wellington in New Zealand ($\mathsf{B}$ for Bob), lying on
	the latitudes $\varphi\approx41^{\circ}$N~and~S, respectively. Alice, an addicted traveller, sets off  from $\mathsf{A}$ and moves eastward along the parallel to some point $\mathsf{C}$. By \mbox{$\lambda\in\lbrack0,2\pi)$} we denote the difference of longitudes (in the sense of \cite[Problem VIII, p. 170]{Kei05}) between $\mathsf{A}$ and
	$\mathsf{C}$, see Fig.~\ref{fig1}.  At this point she tosses  \linebreak

 \noindent
 	a biased   coin to choose her  destination, deciding to either  return to $\mathsf{A}$ or travel to $\mathsf{B}$. 
We assume that the coin's bias is such that her odds  of returning  home are inversely
	proportional to the ratio of squared (Euclidean) distances between $\mathsf{C}$ and the
	potential destinations $\mathsf{A}$ and $\mathsf{B}$. 
	That is, putting  $p$ for the probability of Alice going from~$\mathsf{C}$ to~$\mathsf{A}$, we have $p/(1-p)=|\mathsf{CB}|^2/|\mathsf{CA}|^2$.
	 Hence, Alice has a tendency to go to the place closer to her current location. In fact, it follows easily that 
$|\mathsf{CA}| = 2\sqrt{1-p}$ and 
$|\mathsf{CB}| = 2\sqrt{p}$. 
	To express $p$ in terms of the central angles, one can apply the   \textit{haversine} and
	\textit{havercosine} functions, so appreciated by   navigators of all ages  ($\operatorname{haversin}x:=\sin^{2}\tfrac{x}{2}$, 
	$\operatorname{havercosin}x:=\cos^{2}\tfrac{x}{2}$). 
	From the law of cosines  it follows that   $p=\operatorname{havercosin}(  \theta_{\mathsf{CA}})$  and  $1-p=\operatorname{havercosin}(\theta_{\mathsf{CB}})
	$, where $\theta_{\mathsf{CA}}$ and $\theta_{\mathsf{CB}}$ are the central angles between $\mathsf{C}$ and $\mathsf{A}$, and between $\mathsf{C}$ and $\mathsf{B}$, respectively. 
	 To express $p$ in terms  of geographic   coordinates, we call on the renowned \textit{haversine formula  }\cite{Men1797}, obtaining 
		$p = 1-\cos^{2}\varphi\cdot \operatorname{haversin}\lambda$.

	Should fate send Alice back to $\mathsf{A}$, her trip is complete and she is done with travelling (at least for some time). Assume she finds herself at $\mathsf{B}$. As much as she loves visiting Bob, sooner or later she needs to get back home.
	So one day Alice departs to the east along the parallel, arriving at point $\mathsf{D}$ such that the difference of
	longitudes of $\mathsf{B}$ and $\mathsf{D}$ is again equal to $\lambda$, i.e., $\mathsf{D}$ is antipodal
	to $\mathsf{C}$. However, once at $\mathsf{D}$, she   decides on her destination in the same manner as before; namely, she goes from $\mathsf{D}$ to $\mathsf{A}$ with  probability
	$\operatorname{havercosin}(\theta_{\mathsf{DA}})  = 1-p$, or to $\mathsf{B}$ with 
	probability $\operatorname{havercosin}(  \theta_{\mathsf{DB}})  = p$, where $\theta_{\mathsf{DA}}$ ($=\theta_{\mathsf{CB}}$) and $\theta_{\mathsf{DB}}$ ($=\theta_{\mathsf{CA}}$) are the central angles between $\mathsf{D}$ and $\mathsf{A}$, and between $\mathsf{D}$ and $\mathsf{B}$, respectively. In the latter case, having spent a few extra  days at Bob's place, our vacillating traveller again makes a journey to $\mathsf{D}$, repeating this procedure until   eventually returning  to $\mathsf{A}$.

\begin{figure}[b]
		\vspace{-7mm}
		\includegraphics[scale=0.425]{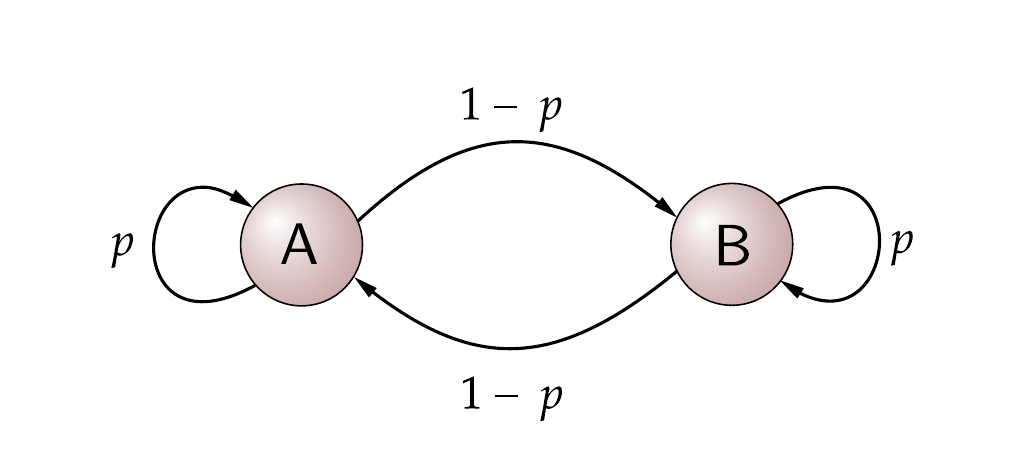}
		\vspace{-3mm}
		\captionsetup{width=0.6\linewidth}
		\caption{
			Alice's travelling as a	symmetric two-state Markov chain. Clearly, the chain is irreducible iff $p <1$.}  \label{fig2}
	\end{figure}

	One can now ask: what is the average number of times Alice will visit Bob
	before getting home? Somewhat surprisingly, the answer depends
	neither on $\varphi$, i.e., the localization of the antipodal cities, nor
	on $\lambda$, and it is always $1$, 
	unless 
	$\varphi=\pm \pi/2$ or $\lambda=0$, in which case all Alice's adventures are imaginary.
	Indeed,  
	generically, we deal here with an irreducible two-state ($\mathsf{A}$ and $\mathsf{B}$) symmetric Markov chain in which the mean return time to $\mathsf{A}$ is equal to $2$, see
	Fig. \ref{fig2}.

	\medskip
	
	From the characters' names one might get the impression that (quantum) information theory is  involved here somehow, and this is indeed the case. Namely, let us replace the globe with the unit
	\textit{Bloch sphere} $S^{2}$, which is isomorphic to $\mathbb{CP}^{1}$ \cite[p. 61]{HeiZim11}. For   $\ket{z} \in \mathbb{CP}^{1}$ we  denote the corresponding element of $S^{2}$ by $r_{z}$. We then have  $|\braket{w|z}|^{2}=\frac{1}{2}(r_{w}\cdot r_{z}+1)$ for $\ket{w}, \ket{z} \in \mathbb{CP}^{1}$, see \cite[p. 63]{HeiZim11}. Next, we swap the antipodal cities $\mathsf{A}$ and $\mathsf{B}$
	for the Bloch vectors $r_{z_{0}},r_{z_{1}}$, where $r_{z_{1}}=\mminus\,r_{z_{0}}$, related to the elements of the orthonormal projective basis $\{\ket{z_{0}}, \ket{z_{1}}\}$ of $\, \mathbb{CP}^{1}$, and travels along parallels for the rotation $O_{\lambda}$ through the angle $\lambda$ about the N-S axis of $S^{2}$. By $U$ we denote the \textit{(projective) unitary operator} corresponding to 
	this rotation via $r_{Uz}=O_{\lambda}(r_{z})$, where $\ket{z} \in \mathbb{CP}^{1}$ \cite[p. 88]{HeiZim11}. Finally, the coin tossing is swapped for the \textit{rank-}$1$ \textit{projection-valued measurement} (PVM) consisting of  $P_{0}, P_{1}$ such that  $P_{0}+P_{1}=\mathbb{I}$ and related to the basis, i.e., 
	$P_{0}\!\ket{z}=\ket{z_{0}}$ and $P_{1}\!\ket{z}=\ket{z_{1}}$ for $\ket{z} \in\mathbb{CP}^{1}$.

	We analyse the situation where successive measurements are performed on a qubit, i.e., on a two-dimensional quantum system, whose evolution between two subsequent
	measurements is governed by $U$. Assume that $\ket{z_{0}}$ is the initial state of the system   and  the instrument describing the measurement process is repeatable.  It follows that the
	probability  $\textrm{P}_{i_{1},\ldots,i_{n}}$  of obtaining a string of measurement  outcomes $(i_{1}%
	,\ldots,i_{n})$, where $i_{m} \in \{0,1\}$ for $m=1,\ldots,n$ and $n\in\mathbb{N}\setminus \{0\}$, is
	given by the celebrated \textit{Wigner formula} \cite{Wig63}:%
	\begin{equation}\label{wignerForm}
 	\textrm{P}_{i_{1},\ldots,i_{n}}=\|  P_{i_{n}}U\cdots P_{i_{1}}U\ket{z_{0}}\|
	^{2}. 
    \end{equation}
It follows that
\begin{equation}\label{markovForm}
	\textrm{P}_{i_{1},\ldots,i_{n}} =  p_{0i_{1}}\prod_{m=1}^{n-1}p_{i_{m}i_{m+1}}, 
\end{equation}
where $p_{jl}:=\|  P_{l}Uz_{j}\|  ^{2}=|   \braket{ z_{l}%
	|U|z_{j} } |  ^{2}$ with $j,l\in \{0,1\}$ is the probability that  we obtain $l$ as the measurement outcome, provided that the preceding measurement yielded~$j$ \cite{SloSzc17, SloZyc94}.  Thus, we have a Markov chain on the set of symbols 0 and 1 with the initial
distribution concentrated at 0 and the doubly  stochastic transition matrix $P:=(p_{jl})_{j,l=0,1}$. 
 Note that the combined evolution of states is also Markovian with two states $\ket{z_{0}}$ and  $\ket{z_{1}}$, the initial
distribution concentrated at $\ket{z_{0}}$, and the same transition matrix $P$.

	We put 
	\vspace{-1mm}
	\begin{equation}\label{andef}
	\vspace{-1mm}
	a_{0}:=1\ \ \textrm{ and } \ \  a_{n}:=\textrm{P}_{\uunderbrace{\scriptstyle1\cdots1}_{\text{$\scriptstyle n$}}} 
	 \ \ \textrm{ for }\ n\in
	\mathbb{N}\setminus \{0\}
	\end{equation}
	and   
	\vspace{-1mm}
	\begin{equation}\label{bndef}
	\vspace{-1mm}
	b_{n}:=\textrm{P}_{{\uunderbrace{\scriptstyle1\cdots1}_{\text{$\scriptstyle n$}}0}} = a_{n}-a_{n+1}  \ \ \textrm{ for }\ n\in\mathbb{N}.
	\end{equation}
	That is, $b_n$ is the probability of obtaining the outcome 0 for the first time in the $(n+1)$-th measurement (i.e., the probability that Alice returns home only after having landed  $n$ times in  \textsf{B}). 
	The mean return time to $\ket{z_{0}}$ (i.e., one plus the average number of visits  Alice pays to Bob) is given by 
	\begin{equation}\label{Mdef}
	M :=\sum_{n=0}^{\infty}(  n+1)  b_{n}.
	\end{equation}
	 Summation by parts (Abel transformation) allows us to express $M$ as follows: 
	\begin{align} \label{MAbel}
	M    =\sum_{n=0}^{\infty}(  n+1)  (a_{n}-a_{n+1})
	= \sum_{n=0}^{\infty} a_{n} - \lim_{N\rightarrow\infty}Na_{N}.
\end{align}

In order to calculate $M$, we first need to determine the transition matrix. Observe that 
\begin{align*}
	p_{00}& =\|  P_{0}U\ket{z_{0}}\|  ^{2}=|  \braket{  z_{0}
		|U|z_{0}} |  ^{2}\\[0.3em]
	&  =\tfrac{1}{2}(r_{z_{0}}\cdot r_{Uz_{0}}+1)=\tfrac{1}{2}(r_{z_{0}}\cdot
	O_{\lambda}(r_{z_{0}})+1) 
	\\[0.3em]
	& = 
	\tfrac{1}{2}(\cos\theta_{\mathsf{CA}}+1)  =\operatorname{havercosin}(  \theta_{\mathsf{CA}})
	=p,
\end{align*}
and so $p_{01}=p_{10}=1-p$ and $p_{11}=p$.   
Therefore, 
$a_n=(1-p)p^{n-1}$ and 
$b_{n}=(1-p)^{2}p^{n-1}$ for $n\! \in \mathbb{N}\setminus\{0\}$, and $b_{0}=p$. Thus, from \eqref{Mdef} we obtain 
	\begin{equation*}
	M  =  p  + \sum_{n=1}^{\infty}(1-p)^{2}np^{n-1} +\sum_{n=1}^{\infty}(1-p)^{2} 
	p^{n-1}.
	\end{equation*}	
 Clearly, if $p=1$, then $M = 1$, and if $p<1$, then 
$M 
	=  (1-p)^{2} (\tfrac{1}{1-p})' + (1-p) +  p 
	 =  2
$. Observe that  $p = 1$ iff  $\operatorname{havercosin}(  \theta_{\mathsf{CA}}) = 1$ iff  
	$\cos^{2}\varphi\cdot \operatorname{haversin}\lambda = 0$  
	iff $\varphi=\pm \pi/2$ or $\lambda=0$, which is in turn equivalent to 
	$U\ket{z_{0}}=\ket{z_{0}}$. 
On the other hand, we have $\lim_{n \to \infty} n a_n = 0$, so it follows from  \eqref{wignerForm} and \eqref{MAbel} that 
	\begin{align*}
	M  = \sum_{n=0}^{\infty}
	\|(P_{1}U)^{n}\ket{z_{0}}\|^{2}.  
	\end{align*}
 In consequence, 
we obtain 
	\begin{equation*}
\sum_{n=0}^{\infty}\|  (  P_{1}U)  ^{n}\ket{z_{0}} \|  ^{2}= \left\{
	\begin{array}{ll}
	1 & \textrm{if }\ U\ket{z_{0}} = \ket{z_{0}},\\
	2 & \textrm{if }\ U\ket{z_{0}} \neq  \ket{z_{0}}\text{.}
	\end{array}
	\right.
	\end{equation*}
	independently on $U$ and $\ket{z_{0}}$.
	Recall that $P_{1}$ is the orthogonal projection on the hyperplane orthogonal to $z_{0}$.
	
	\smallskip
	
	The  primary aim of the present paper is to extend this elementary result to
	higher dimensions. For simplicity, from now on we abandon  the projective approach and stick to Euclidean spaces.  We claim that
	for a~generic choice of $\,U\! \in \mathcal{U}(\Ce^d)$ and $z \in \Ce^d$  such that $||z||=1$  we have  
	\begin{equation}	\label{gencase}
	\vspace{1mm}
	\sum_{n=0}^{\infty}||(PU)^{n}z||^2 = d,\end{equation}
	where $P$ stands for the orthogonal projection in the direction of $z$, i.e., on $\spann\{z\}^\bot$. More specifically, the following theorem holds.
	\begin{theorem}	\label{thmMain}
		 We have
		\vspace{1mm}
		\begin{equation*}
		\vspace{1mm}
		\sum_{n=0}^{\infty}||(PU)^{n}z||^2\,=\,d-\!\sum_{\lambda \in \sigma(U)}\dim(\Theta \cap V_\lambda).
		\end{equation*}
		where $\,\Theta\!:=\!\spann\{z\}^\bot$, $P := \mathbb{I} - \ket{z}\!\bra{z}$, and $\,\{V_\lambda\}_{\lambda \in \sigma(U)}$ is  the family of (orthogonal) ei\-gen\-spa\-ces of $\:U\hspace{-0.5mm}$,  i.e., for $\,\lambda\!  \in\! \sigma(U)$  we put $\,V_\lambda:=\operatorname{Ker}(U-\lambda  \mathbb{I})$.
	
	\end{theorem}

\begin{remark*}
Generically, we have $\sum_{\lambda \in \sigma(U)}\dim(\Theta \cap V_\lambda) = 0$, so \eqref{gencase} follows.
\end{remark*}

 It is straightforward to verify Theorem \ref{thmMain} in the particular case of $z$ being an eigenvector of $U$. In the series at the left-hand side all but the first term (which is equal to $1$) 
vanish, and the sum at the right-hand side gives $d-1$, because there are $d-1$ linearly independent eigenvectors of $U$ in $\Theta$. In general, however, the proof is more demanding, see Sec. \ref{proofs}. 
 
\begin{figure}[h]
 	\vspace{2mm} 
		\includegraphics[width=\linewidth]{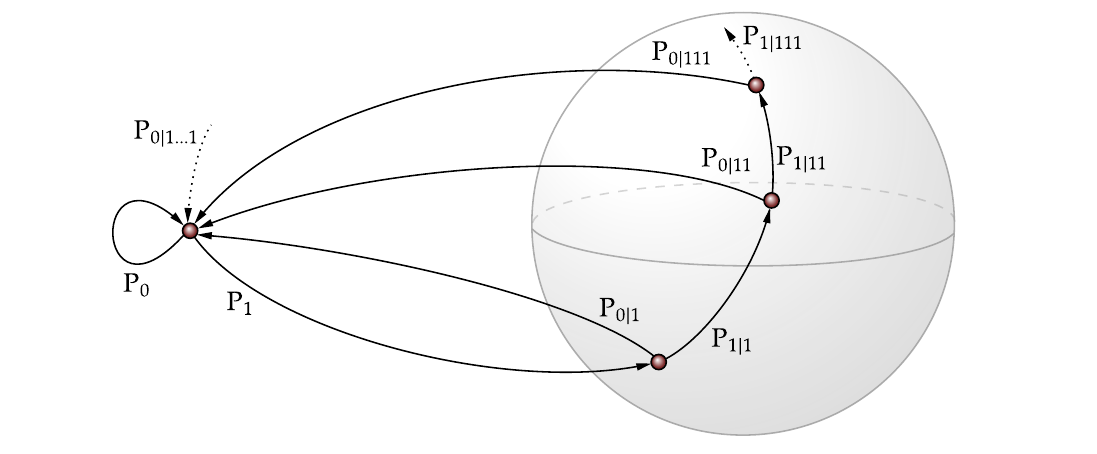}
	 		\vspace{-6mm} 
		\caption{An aggregated Markov chain in the case of $d=3$. By $\mathrm{P}_{i_n|i_1\ldots i_{n-1}}$ we denote the conditional probability of the system outputting $i_n$, provided that so far it has emitted  the outcomes $i_1, \ldots, i_{n-1}$. }  \label{fig3}
	\end{figure}

Theorem \ref{thmMain} has a quantum-mechanical interpretation: in essence, the same as in the qubit case (though geographical analogies are no longer possible). Namely, instead of a~PVM consisting of two rank-1 projections, we now have a PVM comprising one projection of rank $1$ and one projection of rank $d\, \mminus\, 1$, as well as the \mbox{L\"{u}ders instrument} corresponding to this PVM \cite{Busch2009}. In consequence, the two-state Markov chain  is replaced by an aggregated Markov chain 	\cite{Slo03} with two outcomes 0 and 1 and with hidden state space given by the disjoint union of a point, corresponding to $z_0$, and an (at most) countable subset of the projective space $\mathbb{CP}^{d-2}$, which takes the place of $z_1$,
	see Fig. \ref{fig3}. Note that \eqref{wignerForm} and  \eqref{andef} -  \eqref{MAbel} are still valid; however, \eqref{markovForm}  no longer holds as symbolic dynamics is no longer Markovian.

	Accordingly, Theorem~1 provides another operational (physical) meaning 
	to the number of quantum degrees of freedom,\footnote{The authors would like to thank Pawe{\l} Horodecki for suggesting this idea.} i.e., to the dimensionality of
	the 
	Hilbert space underlying the quantum system. 
	The question of how to determine this dimension is not only of theoretical interest but also of utmost practical importance,  because in quantum information theory the system's dimension is regarded as an important resource: in  higher-dimensional spaces more powerful protocols are available. 
	This long-standing problem has been addressed from many perspectives, see, e.g., \cite{Bruetal08,Cai, Galetal10, Hendrych, Maharshi, Vicente, Wehner, Wolf}. Theorem 	\ref{thmMain} offers the possibility of estimating (from below) the dimension of the system from the statistics of a~projective measurement performed on this system. 
	See Section \ref{sec_app} for further discussion.

  In this context, it is noteworthy that convergence in Theorem \ref{thmMain} is geometric.
Namely, let $\rho$ stand the spectral radius of an~operator, i.e., the largest absolute value of its eigenvalues. In Lemma \ref{new2} we will show that $\rho(PU|_{W^\bot \cap\, \Theta})< 1$, where $W\!:=\bigoplus_{\lambda \in \sigma(U)}  (\Theta \cap V_\lambda)$, i.e., $W$ is the maximal subspace of $\Theta$ that is invariant under $U$. Then 
\begin{theorem} \label{thmConv}
	For $n \in \mathbb{N}$ sufficiently large we have  $$||(PU)^{n}z|| < r^n,$$ where $\rho(PU|_{W^\bot \cap\, \Theta}) < r < 1$.
\end{theorem}
\noindent
For the proof of this theorem, see Sec. \ref{proofs}.

	\smallskip
	 
	A result analogous to Theorem \ref{thmMain} holds for real vector spaces,  the formula is slightly more complicated though, because orthogonal matrices need not be diagonalizable.  
	Let  $R$ be an orthogonal operator on $\R^d$. Let $\sigma(R)$    denote its real spectrum; obviously,   $\sigma( {R})  \subset \{\mminus 1, 1\}$.
	For $\lambda \in \sigma( {R})$    put $W_\lambda:=\operatorname{Ker}(R- \lambda \mathbb{I})$, and by
	$A_1, \ldots, A_k$, where $k \leq d/2$, denote  the  \emph{invariant planes} of $R$, i.e., $A_j$, where $j = 1, \ldots, k$, is a two-dimensional subspace of~$\R^d$ with the property that there exists $\varphi_j  \in  \R$ such that with respect to every  orthonormal basis of $A_j$ we have
	$$
	\vspace{2mm}
	R|_{A_j} \sim  \begin{bmatrix}
	\phantom{\mp} \cos \varphi_j & \!\!  \pm\sin \varphi_j \,  \\  \mp \sin \varphi_j &  \phantom{\pm} \!\! \cos \varphi_j\, 
	\end{bmatrix}$$
	with the sign depending on the orientation of the basis. Clearly, $W_{\vphantom{\minus }1}, W_{\minus 1}, A_{\vphantom{\minus }1}, \ldots, A_{\vphantom{\minus } k}$  constitute an orthogonal decomposition of $\R^d$.
	Also,~let  $z \in \R^d$ be a~unit vector. In the real case,  by  $\Theta$ we denote  the orthogonal complement of $z$ in $\R^d$, and  by $P$ the orthogonal projection on $\Theta$.
	
	\begin{theorem}	\label{thmReal}  
		We have
		\vspace{1mm}
		\begin{equation*}
		\sum_{n=0}^{\infty}||(PR)^{n}z||^2=d-\sum_{\lambda\in \sigma(R)} \dim(\Theta \cap W_\lambda)-   2|K|, \end{equation*}
		where  $K:=\{j=1, \ldots, k \colon A_j \subset \Theta\}$.
	\end{theorem}
	
	\noindent
As before, see Sec. \ref{proofs} for the proof. 
		
		Note that $\,2|K|  =  \sum_{j \in K }  \dim A_j  =  \sum_{ j \in K }  \dim(\Theta  \cap A_j)$,
	so the claim of Theorem \ref{thmReal}  can be  rewritten as
	\begin{equation*}
	\sum_{n=0}^{\infty}\|(PR)^{n}z\|^2=d\: - \!\sum_{\lambda\in \sigma(R)} \dim(\Theta \cap W_\lambda) -  \sum_{ j \in K }  \dim(\Theta  \cap A_j). \end{equation*}
Note also that in the generic case we have $$\sum_{n=0}^{\infty}||(PR)^{n}z||^2=d.$$ 
 Let us illustrate Theorem \ref{thmReal} with the following example.
	\begin{example*}
		We fix a unit vector $z\in \R^2$ and let $P$ stand for the orthogonal projection on $\Theta:=\spann\{z\}^\bot$. 
		Under the standard identification of $\R^2$ with $\Ce$ we have 
		$z = \e^{\I\theta}$ for some $\theta \in \R$, so $\Theta=\e^{\I(\theta+\pi/2)}\R$.  
		It is straightforward to verify that  
		$P$ acts as 
		$$P\colon  \Ce \ni r{\hspace{-0.01mm}\e}^{\I \kappa} \mapsto r\sin( \kappa -\theta ) \e^{\I( \theta + \pi / 2)} \in \Ce,$$ where $r \geq0$ and $\kappa \in \R$. 
		We investigate $S:= \sum_{n=0}^{\infty}||(PR)^{n}z||^2$ with $R$  assumed to be an orthogonal operator on $\R^2$, i.e., a rotation or reflection.

		Firstly, let $\varphi \in [0, 2\pi)$. Consider $Rw := \e^{\I\varphi}\!w$ for $w \in \Ce$, i.e., $R$ is the rotation about the    origin through  the angle $\varphi$. For~$n \geq 1$ we   easily obtain 
	 $$
		(PR)^{n}z=  \sin\varphi\,
		\cos^{n-1}\!\varphi \, \e^{\I(  \theta+\pi/2)  };$$  therefore, 
		$S=1+  \sum
		_{n=0}^{\infty}\sin^{2}\varphi \cos^{2n}\varphi$.  
		
		\begin{itemize}[leftmargin=5mm]
			\itemsep=0.5mm
			\item If $\,\varphi \notin \{0, \pi\}$, then $S=2$.
			Note that $\sigma(R)= \varnothing\,$ and  $A_1=\R^2$, so $\dim(\Theta \cap A_1)=1$ and $K=\varnothing$.

			\item 
			If $\,\varphi=0$, i.e., $R$ is the identity,
			then $PRz=Pz=0$ and $S=1$. Clearly,  $\sigma(R)=\{  1\}$, $W_{1}=\R^2$, and $\dim(  \Theta\cap W_{1})=1$. 
			
			\item If $\,\varphi=\pi$, i.e., $R$ is the    point reflection $w\mapsto\mminus w$ through  the   origin, then  $PRz= \mminus Pz=0$ and  $S=1$. We~have 
			$\sigma(R)=\{\mminus1\}$, $W_{\minus 1}=\R^2$, and $\dim(  \Theta\cap W_{\minus 1})=1$.
			
		\end{itemize}

		Next, consider $Rw := \e^{2\I\varphi}\overline{w}$ for $\varphi \in [0,\pi)$, i.e., $R$ is the reflection about the line through the origin which makes an angle $\varphi$ with the real axis. Clearly, we have $\sigma(R)=\{ \mminus 1,1\}  $, and  the eigenspaces of $R$ read $W_{1}=\e^{\I\varphi}\R$,   $\,W_{\minus 1}=\e^{\I(\varphi+\pi/2)}\R$.
		It follows that 
		$$
		(PR)^{n}z=- \sin(2(\varphi - \theta) )
		\cos^{n-1}(2(\varphi-\theta))  \e^{\I(  \theta+\pi/2)}$$
		for $n\geq 1$. In consequence,  
		$S= 1+ \sum_{n=0}^{\infty}\sin^{2}(2(\varphi-\theta))\cos^{2n}(2(\varphi-\theta))$.
		\begin{itemize}[leftmargin=5mm]
			\itemsep=0.5mm
			\item 
			If $\,\varphi  \notin \{\theta, \theta+\pi/2\}$, i.e., $z \notin W_{1} \cup W_{\minus 1}$, then  
			$S=2$. Note that   $\dim(  \Theta\cap W_{1})  =\dim(
			\Theta\cap W_{\minus1})=0$.
			\item 
			If $\,\varphi=\theta$, i.e., $z \in W_{1}$ and $\Theta=W_{\minus 1}$, then $PRz\!=\!Pz\!=\!0$ and $S\!=\!1$. Obviously,  $\dim(  \Theta\cap W_{ 1})\!=\!0$  and $\dim(\Theta\cap W_{\minus1})\!=\!1$.
			\item 
			If $\,\varphi=\theta +\pi/2$, i.e., $z\! \in\! W_{\minus 1}$ and $\Theta=W_{1}$, then $PRz\!=\!\mminus Pz\!=\!0$ and $S\!=\!1$. We have   $\dim(  \Theta\cap W_{ 1}) \! =\!1$  and  $\dim(  \Theta\cap W_{\minus1}) \! =\!0$.
			
		\end{itemize}
		
		\vspace{2mm}
		
	\end{example*}

	\section{Proofs}\label{proofs}
	 Adopting the standard convention that raising any operator to the null power yields the identity,  we see that both theorems are 
	hyper-obvious for $d=1$. Thus, in order to avoid trivial statements, we assume that $d\geq 2$. For the whole of this section we adopt the following notation. 
As before, the orthogonal complement of $z$ in $\Ce^d$ (or, in the proof of Theorem~3, in $\R^d$) is denoted by   $\Theta$,  the orthogonal projection on $\Theta$ by $P$, and  $W\!:=\bigoplus_{\lambda \in \sigma(U)}  (\Theta \cap V_\lambda)$ is the maximal subspace  of $\Theta$ invariant under $U$. By $||\cdot||$ we  denote the spectral norm on $\mathcal{L}(\Ce^d)$, i.e., the operator norm induced on the space of linear transformations of $\,\Ce^d$ by the Euclidean norm, while $\rho$ stands for the spectral radius of an operator.

	\begin{lemma}\label{evU}
		
		If  $\,v \in \Theta$	 is an eigenvector of $\,PU$ with eigenvalue $\mu \in \Ce$ and $|\mu|=1$, then
		$v$	 is  an eigenvector of~$\,U$~with eigenvalue $\mu$.  
	\end{lemma}
	
	\begin{proof}
		  Fix $v \in \Theta  \setminus \! \{0\}$ and $\mu \in \Ce$, $|\mu|=1$, such that  $PUv=\mu v$.  Then $||PUv||=||v||$.  Obviously, the unitarity of $U$ gives $||Uv||=||v||$. Moreover,   $ ||Uv||^2=||PUv||^2+|\braket{z| Uv}|^2$. It follows that   $\braket{z|Uv}=0$, i.e., $Uv \in \Theta$. Hence, $Uv=PUv=\mu v$, as required. 
	\end{proof}

	\begin{lemma} \label{new22}
		We have $\,PU(W^\bot) \subset W^\bot \cap \Theta$.
	\end{lemma}
	\begin{proof}
		Clearly, as $P$ is a projection on $\Theta$, it is sufficient to show that $W^\bot$ is invariant under $PU$.	
		Letting $u \in W^\bot$,  we obtain $Uu \in W^\bot$ due to the invariance of $W^\bot$ under $U$.  Thus,   
		$\braket{PUu|w}=\braket{Uu|Pw}=\braket{Uu|w}=0$ 	for every $w \in W$, and so $PUu \in W^\bot$. 	 
	\end{proof}

	\begin{lemma}\label{new2}
		$PU$ is an~endomorphism on $\:W^\bot \cap \Theta\,$ and
		$\rho(PU|_{W^\bot \cap\, \Theta})< 1$.

	\end{lemma}
	\begin{proof}
		\vspace{-1mm}
		The fact that $PU$ is an~endomorphism on $W^\bot \cap\: \Theta$ follows easily  from Lemma~\ref{new22}. As for the spectral radius,   observe   that $\rho(PU) \leq  1$ as    $\rho(PU)  \leq ||PU||   \leq  ||P||   \cdot    ||U||$ and \mbox{$||P|| = ||U|| = 1$}.
		Consequently,  $\rho(PU|_{W^\bot \cap\, \Theta})\leq 1$.
		Let $\mu \in \sigma(PU|_{W^\bot \cap\, \Theta})$ and let      $v \in {W^\bot \cap\, \Theta}$ be    an eigenvector of $PU$ with eigenvalue $\mu$.
		In particular, we have  $v \notin W$, which implies that $v \notin \Theta \cap V_\lambda$ for  every $\lambda \in \sigma(U)$. Hence,  $v$ is not an eigenvector of $U$ and  Lemma~\ref{evU} assures that $|\mu|<1$, as desired.  		 
	\end{proof}

The preceding three lemmas pave  the way for the following result, which not only is a crucial step in the proof of Theorem 	\ref{thmMain}, but also plays a key role in studying the symbolic dynamics generated by the quantum system under consideration \cite[Sec. 1.3]{phd}.

	\begin{lemma}\label{lemma4} We have
		\vspace{-1mm}
	\begin{equation*}
	 \lim\limits_{n \to \infty}\tr((PU)^{n} (PU)^{* n}) =\sum_{\lambda \in \sigma(U)}\dim(\Theta \cap V_\lambda).
	\end{equation*}
\end{lemma}	
	
	\begin{proof} 
	
Clearly, $\dim W = \sum_{\lambda \in \sigma(U)}\dim(\Theta \cap V_\lambda)$, so the above claim can be rewritten as
\begin{equation*}  \lim\limits_{n \to \infty}\alpha_n =\dim W,
\end{equation*}
where $\alpha_k:=\tr((PU)^{k} (PU)^{* k})$ for $k \in \N$. 
 First, recall that $W$ is invariant under $U$. We observe  that $PU|_{W}=U|_W$. Indeed, for $w \in W$ we obtain  $Uw \in W \subset \Theta$, and thus $PUw=Uw$. In consequence, $PU|_{W}$ is unitary, and so  
\begin{equation}\label{PUW} 
	||PUw||=||w|| \ \ \textrm{ for every } \ w \in W.
\end{equation}
 Next,  from Lemma \ref{new2} it follows that  $\left(PU|_{W^\bot \cap\, \Theta}\right)^n \to 0$ as $n \to \infty$,  so, via Lemma \ref{new22}, we obtain 
\begin{equation}\label{PUWbot} (PU)^n u =(PU)^{n-1} (PUu) \xrightarrow{n \to \infty} 
	0 \ \ \textrm{ for every } \ u \in W^\bot.
\end{equation}

 Now, we put $\tilde{d}:=\dim W$ and choose an orthonormal basis  $\{w_1, \ldots, w_d\}$   of $\,\Ce^d\,$ such that $W=\spann\!\left\{w_{\vphantom{\tilde{d}}1}, \ldots, w_{\tilde{d}}\right\}$ and  $W^\bot=\spann\! \left\{w_{\tilde{d}+1}, \ldots, w_{\vphantom{\tilde{d}}d}\right\}$.
It follows that
\begin{align*}
	\alpha_n
	&	= \tr ( (PU)^{*n}(PU)^n)
	=\sum_{i=1}^{d} \braket{w_i|(PU)^{*n} (PU)^nw_i}  
	=\sum_{i=1}^{d}|| (PU)^nw_i||^2.
\end{align*}
Applying \eqref{PUW} and \eqref{PUWbot}, we obtain
$$\lim\limits_{n \to \infty} || (PU)^nw_i||=\left\lbrace\begin{array}{ll}
	1 & \textrm{if }\: i = 1, \dots,  \tilde{d}  
	\\[0.2em]
	0 	& \textrm{if }\:  i = \tilde{d}+1, \dots,  {d} 
\end{array}\right.
$$
which gives $\lim\limits_{n \to \infty}\alpha_n=\tilde{d}\,$ and concludes the proof.	
	\end{proof}

	\begin{proof}[\textbf{Proof$\,$ of$\,$ Theorem 	\ref{thmMain}}] 
		 Let us    show   that for every $n\in \N$ we have
		\begin{equation*}
		||(PU)^{n}z||^2=\alpha_n-\alpha_{n+1},
		\end{equation*}
		where, as before,  $\alpha_k$ stands for $\tr((PU)^{k} (PU)^{* k})$, $k \in \N$. Put $P_z:=\mathbb{I}-P$, i.e.,  $P_z$ is the orthogonal projection on $\spann\{z\}$. It follows that 
		\begin{align*}
		||(PU)^{n}z||^2& =
		\braket{(PU)^{n}z| (PU)^{n}z}
		\\ 
		& =
		\tr(   (PU)^{n} P_z (PU)^{* n} )
		\\ 
		& =\tr(  (PU)^{n}(\mathbb{I}-P)  (PU)^{* n})
		\\ 
		&
		=\alpha_n-\tr( (PU)^{n } P  (PU)^{*n })
		\\ 
		&
		=\alpha_n-\tr( (PU)^{n } P UU^*P ^*  (PU)^{*n })
		\\ 
		&
		=\alpha_n-\alpha_{n+1},
		\end{align*} 	
		where $n \in \N$, as desired. 
		As a consequence, we have  
$$		\label{teza}\sum_{n=0}^{\infty}||(PU)^{n}z||^2 
		=  \lim_{N \to \infty }\sum_{n=0}^{N-1}(\alpha_n-\alpha_{n+1}) 
 = 		\tr\mathbb{I} - \lim\limits_{N \to \infty}\alpha_N.$$ 
	To conclude the proof, it suffices to apply Lemma \ref{lemma4}.
	\end{proof}

		\begin{proof}[\textbf{Proof$\,$ of$\,$ Theorem \ref{thmConv}}] 
	Recall from Lemma \ref{new2} that $\rho(PU|_{W^\bot \cap\, \Theta})<1$ and consider $r \in \R$ such that $\rho(PU|_{W^\bot \cap\, \Theta})< r <1$. 
  The celebrated Gelfand's formula implies that 
 $$||(PU|_{W^\bot \cap\, \Theta})^n||^{\tfrac 1n} < r $$  for $n \in \N$ sufficiently large. 
Since $z \in W^\bot$,   Lemma \ref{new22} implies that  $(PU)^n z \in \Theta \cap  W^\bot$ for every $n \in \N\setminus \{0\}$.  In consequence, $$||(PU)^n z|| \leq  ||(PU|_{W^\bot \cap\, \Theta})^n|| < r^n,$$  as desired.
	    \end{proof}
    
	\begin{proof}[\textbf{Proof$\,$ of$\,$ Theorem \ref{thmReal}}] 
	The first step is to use complexification (see,~e.g., \cite[p. 282]{Gant}) and to apply Theorem 	\ref{thmMain}. Let $\tilde{R} \in \mathcal{L}(\Ce^d)$ be
	 the complexification of $R$, i.e.,     $\tilde{R} (x+\I y)=Rx+\I Ry$ for \mbox{$x,y \in \R^d$}.
		We  have $\tilde{R} \in \mathcal{U}(\Ce^d)$,  $\sigma(\tilde{R}) \cap \R = \sigma(R)$ and  $ \sigma(\tilde{R}) \! \setminus  \R  = \{\e^{\pm \I \varphi_1}, \ldots, \e^{\pm \I \varphi_k}\}$, where $\varphi_j \in \R$ for $j \in \{1, \ldots, k\}$, and $k\leq d/2$. For   $\lambda \in \sigma(\tilde{R})$   we put~$V_\lambda:=\operatorname{Ker}(\tilde{R}-\lambda  \mathbb{I})$, and 
 $\tilde{\Theta}$  denotes the orthogonal complement of~$z$ in $\Ce^d$.
		Clearly, for $x,y \in \R^d$ we have
		\vspace{1mm}
		\begin{equation}
		\vspace{1mm}
		\label{Tquiv} x+ \I y \in \tilde{\Theta} \ \Longleftrightarrow\  x - \I y \in \tilde{\Theta}\ \Longleftrightarrow\ x, y \in \Theta;\end{equation}
		in particular, $\tilde{\Theta} \cap \R^d = \Theta$. 
		Also, we put $\tilde{P} \in \mathcal{L}(\Ce^d)$ for  the orthogonal projection on $\tilde{\Theta}$.
		Note that $\tilde{P}$ is the complexification of  $P$  and   $\tilde{P}|_{\R^d}=P$.
		From Theorem 	\ref{thmMain} we obtain 
		\begin{equation*}
		\sum_{n=0}^{\infty}||(\tilde{P}\tilde{R})^{n}z||^2=d-\sum_{\lambda \in \sigma(\tilde{R}) }\dim_\Ce(\tilde{\Theta}_\lambda),\end{equation*}
		where    $\tilde{\Theta}_\lambda:=\tilde{\Theta} \cap V_\lambda$ for $\lambda \in \sigma(\tilde{R})$. 
		Obviously, since $\tilde{P}\tilde{R}|_{\R^n}= {P}{R}$, it follows that  $$	\sum_{n=0}^{\infty}||(\tilde{P}\tilde{R})^{n}z||^2=	\sum_{n=0}^{\infty}||(PR)^{n}z||^2.$$
		Therefore, it suffices to prove that
		\vspace{1mm}
		\begin{equation}
		\label{xx}
		\sum_{\lambda \in \sigma(\tilde{R}) }\dim_\Ce(\tilde{\Theta}_\lambda) = \sum_{\lambda\in \sigma(R)} \dim_\R(\Theta \cap W_\lambda)+    2| K|.\end{equation}
		The proof of \eqref{xx} consists of two parts, addressing separately the real and non-real eigenvalues of $\tilde{R}$.
		
		First,  consider $\lambda \in \sigma({R})$. Note that for $x,y \in \R^d$ we have 
		\begin{equation}
		\label{Requiv}x+ \I y \in V_\lambda \ \Longleftrightarrow \ x, y \in W_\lambda.\end{equation} We now show that 
		\begin{equation}
		\vspace{1mm}
		\label{twoeq1}\dim_\Ce(\tilde{\Theta}_\lambda)=\dim_\R(\Theta \cap W_{\lambda}).\end{equation}
		Let $\{w_i\}_{i =1}^m \subset \R^d$ be a basis of $\Theta \cap W_{\lambda}$  over $\R$. 
		We claim that  those vectors constitute   a basis of $ \tilde{\Theta}_\lambda $ over~$\Ce$. 
		Since real vectors that are linearly independent over $\R$ are linearly independent over $\Ce$ as well, we only need to show that  
		$	\spann_\Ce\{w_i\}_{i =1}^m = \tilde{\Theta}_\lambda.$
		By \eqref{Tquiv} and \eqref{Requiv} we  obtain, respectively, $\Theta \subset  \tilde{\Theta}$ and $W_\lambda \subset V_\lambda$, so $w_i \in  \Theta \cap W_\lambda \subset  \tilde{\Theta}_\lambda$ for each $i=1, \ldots, m$, thus also  $\spann_\Ce\{w_i\}_{i =1}^m \subset \tilde{\Theta}_\lambda$. To verify  that   the opposite inclusion also holds,  consider  $u \in \tilde{\Theta}_\lambda$. Again from  \eqref{Tquiv} and \eqref{Requiv} it follows that  $$\operatorname{Re}u, \operatorname{Im}u \in  \Theta \cap W_{\lambda}  = \spann_\R\{w_i\}_{i =1}^m,$$ from which we easily deduce that $u \in\spann_\Ce\{w_i\}_{i=1}^m$, as claimed. Hence,   \eqref{twoeq1} holds. 
		Summing \eqref{twoeq1} over  $\sigma(R)$ gives
		\vspace{1mm}
		\begin{equation}
		\label{twoeq2}  \sum_{\lambda \in \sigma(\tilde{R})  \cap   \R }\!\!\dim_\Ce(\tilde{\Theta}_\lambda)=\sum_{\lambda\in \sigma(R)} \!\dim_\R(\Theta \cap\, W_\lambda).\end{equation}

		Secondly,  consider $\lambda \in \sigma(\tilde{R})\! \setminus \! \R$.  
		Recall that $V_\lambda$ and $V_{\bar{\lambda}}$ are orthogonal subspaces of $\Ce^d$. Let $x,y \in \R^d$ and observe that   $x+ \I y \in V_\lambda  $ iff $x- \I y \in V_{\bar{\lambda}}$, and also  that  $\braket{x+ \I y|x- \I y}=||x||^2-||y||^2-2\I\!\braket{x|y}$ as well as  $\,||x+ \I y||^2=||x||^2+||y||^2$. We obtain  
		\begin{equation}
		\label{Vquiv}
		x+ \I y  \in \!V_\lambda \Leftrightarrow \braket{x|y}\!=\!0,\  ||x||\!=\!||y||\!=\!\tfrac{1}{\sqrt{2}}||x+ \I y||,  \textrm{ and }  x, y  \in\! A_s  \textrm{ for some }    s   \in \!I_\lambda, \end{equation}
		where $I_\lambda := \{j=1, \ldots, k \  \colon \e^{\pm \I \varphi_j}  =   \lambda  \}$. 	
		We put  $K_\lambda:=\{j \in I_\lambda \colon A_j \subset \Theta \}$ and 
		prove that
		\begin{equation}
		\label{theeeq}
		\dim_\Ce (\tilde{\Theta}_\lambda)  =  |K_\lambda|.
		\end{equation}
		For each $j\! \in\! K_\lambda$ we choose  an orthonormal  basis $\{x_j, y_j\}$ of $A_j$. As the invariant planes of $R$ are mutually orthogonal, $\{x_j, y_j\}_{j\in K_\lambda}$ constitutes  an orthonormal  basis of  $\:\bigoplus_{j\in K_\lambda}\! A_j$. We claim that $\{x_j+\I y_j\}_{j \in K_\lambda}$ is then a basis of $\tilde{\Theta}_\lambda$. We   easily verify that those vectors  form an  orthogonal (so linearly independent) set in $\Ce^d$. Let us show that 
		they generate  $\tilde{\Theta}_\lambda$. For brevity, put $Q_\lambda:=\spann_\Ce\{x_j+\I y_j\}_{j\in K_\lambda}$.

		By \eqref{Tquiv} \& \eqref{Vquiv} we   obtain $x_j+\I y_j \in \tilde{\Theta}_\lambda$ for each $j\in K_\lambda$, thus also $Q_\lambda \subset \tilde{\Theta}_\lambda$. 
		To see that the other inclusion holds as well,   let $v \in  \tilde{\Theta}_\lambda \!\setminus \! \{0\}$ and put  $x:=\operatorname{Re}v$, $y:=\operatorname{Im}v$. 
		It follows  from \eqref{Vquiv} that there exists $s \in I_\lambda$ such that $\{\alpha x,\alpha y\}$ is an orthonormal basis of $A_{s}$, where $\alpha:={ \sqrt{2}}/{||v||}$ provides normalization. Since \eqref{Tquiv} assures that   $x,y \in \Theta$, we deduce that  $s \in K_\lambda$. 
		As the  transition matrix from $\{\alpha x, \alpha y\}$ to $\{x_{s}, y_{s}\}$ is orthogonal,   we have 
		$\alpha x = x_{s} \cos \psi  \mp  y_{s}\sin \psi $ and 
$\alpha y =x_{s}  \sin \psi \pm  y_{s} \cos \psi$ 
		for some  $\psi \in \R$. Hence,  $x+\I y = \alpha^{\minus 1} \e^{\I \psi}(x_{s}\pm\I y_{s})$. If $v\in \spann_\Ce\{x_{s}-\I y_{s}\} \subset V_{\bar{\lambda}}$, then $v \in V_{\lambda} \cap V_{\bar{\lambda}} = \{0\}$, which contradicts the assumption  $v \neq 0$. 
		If $v\in \spann_\Ce\{x_{s}+\I y_{s}\}$, then  obviously $v \in Q_\lambda$. Thus,  $\{x_j+\I y_j\}_{j \in K_\lambda}$ is indeed a basis of $\tilde{\Theta}_\lambda$, and so
		\eqref{theeeq} holds.

		Recall that $K\!:= \{j = 1, \ldots, k \colon A_j \subset \Theta\}$. Clearly, we have $K\! = \bigcup \{K_\lambda \colon \lambda \in \sigma(\tilde{R})\!\setminus \R\}$.
		Note also that $K_{{\lambda}\vphantom{\bar{\lambda}}}\!=K_{\bar{\lambda}}\:$   and $K_{\lambda} \cap K_{\mu} = \varnothing$ for $\mu \notin \{\lambda,  \bar{\lambda}\}$.
		Therefore, summing \eqref{theeeq} over  $\sigma(\tilde{R})  \setminus  \R$, we obtain 
		\begin{equation}
		\label{twoeqq}  \sum_{\lambda \in \sigma(\tilde{R})\setminus \R }\!\!\dim_\Ce(\tilde{\Theta}_\lambda)  = 2 |K|.
		\end{equation}
		Adding   \eqref{twoeqq} to \eqref{twoeq2}  results in  \eqref{xx} and concludes the proof of Theorem \ref{thmReal}. 
	\end{proof}

	\section{Algorithm}\label{sec_app}
	
	As we mentioned above, Theorem 	\ref{thmMain} can be used to estimate the dimension $d$ of the Hilbert space underlying a quantum system. 
   Consider a yes-no measurement (elementary test) represented by a PVM $\Pi = \{\ket{z}\!\bra{z},\, \mathbb{I}-\ket{z}\!\bra{z}\}$, where  {$z$ is a unit vector from $\mathbb{C}^d$}, along with the corresponding \mbox{L\"{u}ders instrument}. 
One can think of applying $\Pi$ as posing the question whether the system is in state $\ket{z}$ or not \cite{PhysRevApplied}. 
This measurement is performed repeatedly in an isochronous manner and between each two subsequent measurements the system undergoes deterministic time evolution governed by a unitary operator $U$. An example of such a system (for $d=3$) is a spin-1 particle subject to a magnetic field rotating the spin, with the measurement answering the yes-no question whether the square of the spin component  {along a  given  axis is zero}  \cite{CruWie10}. 

 Assume that the initial state of the system is $\ket{z}\!\bra{z}$. 
 In the current context,  \eqref{wignerForm}	reads
\begin{equation}\label{PUn}
\textrm{P}_{\uunderbrace{\scriptstyle1\cdots1}_{\text{$\scriptstyle n$}}} =\| (PU)^n {z}\|^{2},
\end{equation} 
where  $P := \mathbb{I}-\ket{z}\!\bra{z}$   and 
	$n \in \N\setminus \{0\}$. 
 Hence,  from \eqref{gencase} we obtain the following   direct formula  
\begin{equation}\label{P1111} 
1 + \sum_{n = 1}^\infty 	\textrm{P}_{\uunderbrace{\scriptstyle1\cdots1}_{\text{$\scriptstyle n$}}}  = d.\end{equation} 
 This opens the way to determining the dimension of the underlying Hilbert space  from the joint probabilities of the measurement results, which can be estimated from repeated runs of the experiment. 
Clearly, the reliability of the resulting dimension witness depends on the device being a POVM and  consisting of two projections, one of them  one-di\-men\-sio\-nal. 

In practice, we  observe the sequence of partial sums $S_N(U) := \sum_{n=0}^{N-1}||(PU)^nz||^2$ that tends  to $d$ 
at least geometrically as 	$r^{2N}/(1-r^2)$ 
with $N \to \infty$, where $r$ is lower bounded by the spectral radius  of $PU|_{W^\bot \cap\, \Theta}$, see Theorem \ref{thmConv}. 
 If an eigenspace of $U$ intersects $\Theta$ non-trivially or it is very close to $\Theta$, then, respectively, $S_N(U)$ does not converge to $d$ or the convergence is very slow as $\rho(PU|_{W^\bot \cap\, \Theta})$ is very close to one.  As a way to circumvent this problem  we propose to consider several different unitaries simultaneously, see 
\hyperref[app]{Appendix}. Let us also point out that the sequence of ceilings $\lceil S_N(U) \rceil$ of the partial sums converges to $d$ even faster than the original series, eventually hitting $d$ for some $N$. 
	
 Actually, the joint probabilities required to find $d$ via  \eqref{P1111} can be inferred from a single sequence of outcomes. Namely, the outcome 0 identifies the underlying (hidden) quantum state as $\ket{z}$, i.e., as the initial state of the  measurement protocol. Hence, whenever   0 appears in the sequence of outcomes, the system is reset to the initial setting. 

 Alternatively, $d$ can be computed as the mean return time to $\ket{z}$. To see this, combine \mbox{\eqref{MAbel} \& \eqref{PUn}}, and invoke   Theorem \ref{thmConv} to verify that the limit in \eqref{MAbel} vanishes. The mean return time to $\ket{z}$ can be estimated from a single sequence of measurement  outcomes by the Monte Carlo method as the average distance between the consecutive occurrences of 0, or, equivalently, as one plus the average length of a series of 1's.
Namely, let $1 \leq {j_1}<{j_2}< \ldots$ stand for the positions in the sequence of outcomes occupied by 0's. Put $T_k:=j_{k}-j_{k-1}$ for the distance between the $(k-1)$-th and $k$-th occurrence of 0, and $T_1:=j_1$:
$$ 
\uunderbrace{ 1, \ldots, 1, \overarrow[0]{${j_1}$}}_{T_1},  \uunderbrace{ 1, \ldots, 1, \overarrow[0]{${j_2}$}}_{T_2},     \ldots \ldots , \overarrow[0]{${j_{k-1}}$}, \uunderbrace{ 1, \ldots, 1, \overarrow[0]{${j_k}$}}_{T_k}, \ldots 
$$
By the strong law of large numbers we get  
$$  
\lim_{N \to \infty} \frac 1 N \sum_{k = 1}^N T_k = d,
$$
almost certainly, as desired.

	\section*{Acknowledgments}
	The authors acknowledge financial support from  the Polish National Science Centre \linebreak under Project No. DEC-2015/18/A/ST2/00274. The authors express their  gratitude to  the anonymous reviewers for valuable comments and suggestions resulting in significant improvements to the manuscript, in particular in clarifying the role and properties of the dimension witness.

	  	\setlength{\bibitemsep}{0.1\baselineskip}       	\printbibliography

  \section*{Appendix}\label{app}

	To determine the system's dimension $d$ more efficiently, we can take $M$ unitary matrices $U_1, \ldots, U_M$ and observe the evolution of the distribution of $d_N(U_1), \ldots, d_N(U_M)$ for successive values of $N$, where $d_N(U) := \lceil S_N(U) \rceil$  is the estimate from below for $d$. In this way we lower the risk of dealing with a slowly convergent series. Nevertheless, there remains the problem of establishing a stopping criterion for this estimation procedure. 
	
\begin{figure}[h]
		\includegraphics[width=0.9625\linewidth]{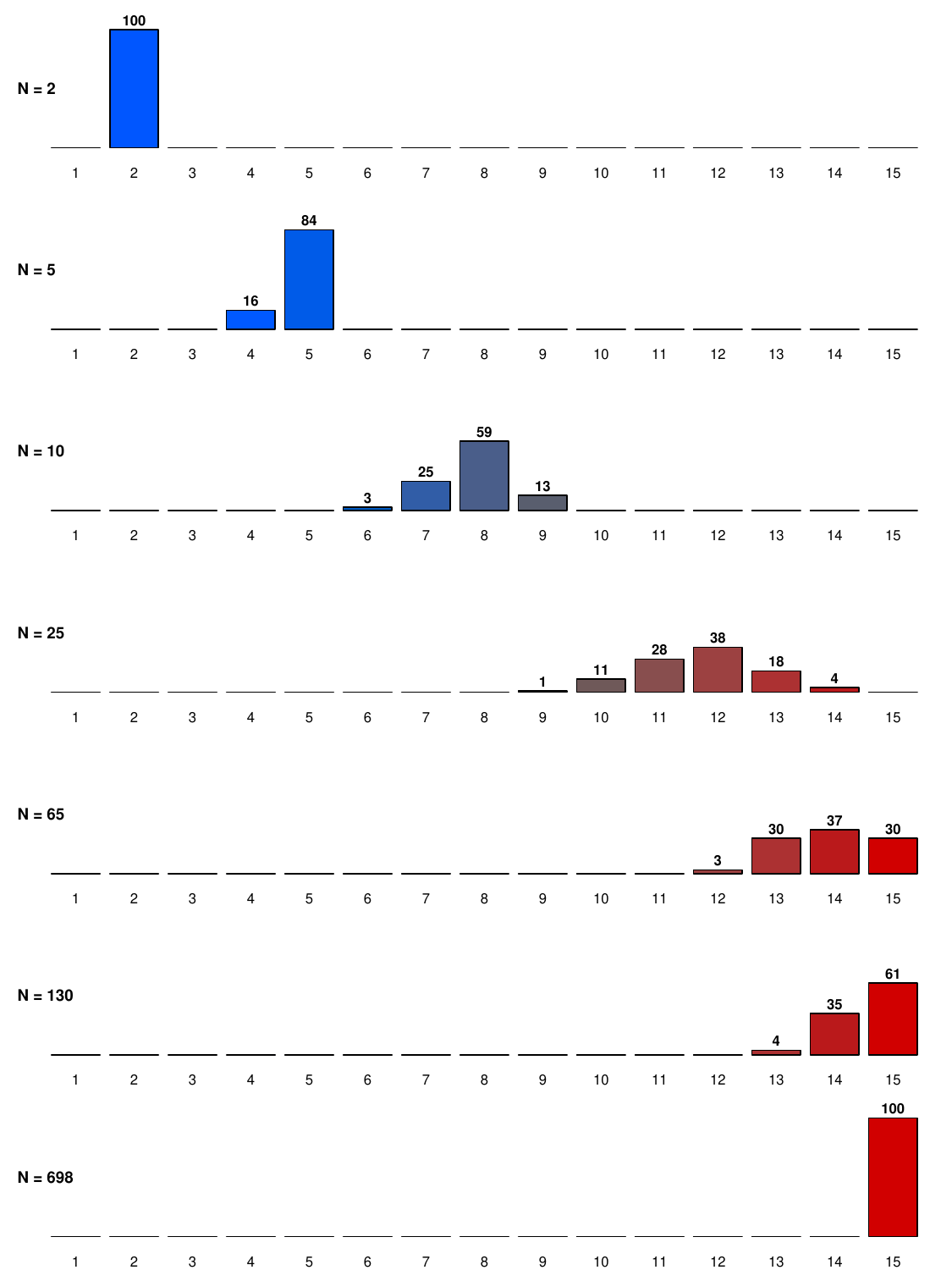}
		
		\vspace{5mm}
		
		\captionsetup{width=0.8\linewidth}\caption{Barplots illustrating the distribution of $d_N(U_1), \ldots, d_N(U_{100})$ for several   values of $N$. The actual dimension of this system is $d=15$.}
		\label{fig_evol}

	\end{figure}

To illustrate this problem, we took $100$ unitary matrices in dimension $d=15$ generated from the Haar distribution (CUE). It turns out that $N = 698$ steps had to be executed  in order for all these unitary matrices to point to the actual dimension  of the system, see Fig.~\ref{fig_evol}. However, one can  argue that the correct result could have been identified much earlier from the shape of these distributions. 
	
In this vein, we propose to take $\tilde{d}:=\max_{i=1, \ldots, M}\{d_{\tilde{N}}(U_i)\}$ 
	as the estimate of $d$ if the following two conditions hold:
	\begin{enumerate}
 	\itemsep=0.5mm
		\item[(i)]
		$\max_{i=1, \ldots, M} \{d_{N}(U_i)\}=\tilde{d}\:$ for $\,N = \tilde{N}-s, \ldots, \tilde{N}-1$,
		\item[(ii)] $
		|\lbrace i = 1, \dots, M \colon d_{\tilde{N}-s}(U_i) = \tilde{d}\, \rbrace | \geq  \beta M\,$,
	\end{enumerate}
	where $\beta \in [0,1]$ and $s \in \mathbb{N}\setminus \{0\}$ are parameters. That is, in terms of the barplots, the far right bar is required  to remain stable (not to move further right) for $s+1$ consecutive steps and to contain at least $\beta M$ of all observations.
	Clearly, this stopping criterion is always met since for each $i=1, \ldots, M$ the sequence $\{d_N(U_i)\}_{N =1 }^\infty$ is non-decreasing and equal to $d$ from some $N$ onwards, and  $\tilde{N}$ is the number of executed steps.
	
		Obviously, there is a trade-off between accuracy and time-efficiency. The parameters $M$, $\beta$, $s$ can be used to find a balance between increasing the probability of the algorithm returning the correct estimate of the system's dimension (by increasing the parameters) and decreasing the number of executed steps (by decreasing the parameters).

\begin{figure}	[h]	 
		
		\vspace{3mm}	
		
 	\includegraphics[width=0.95\linewidth]{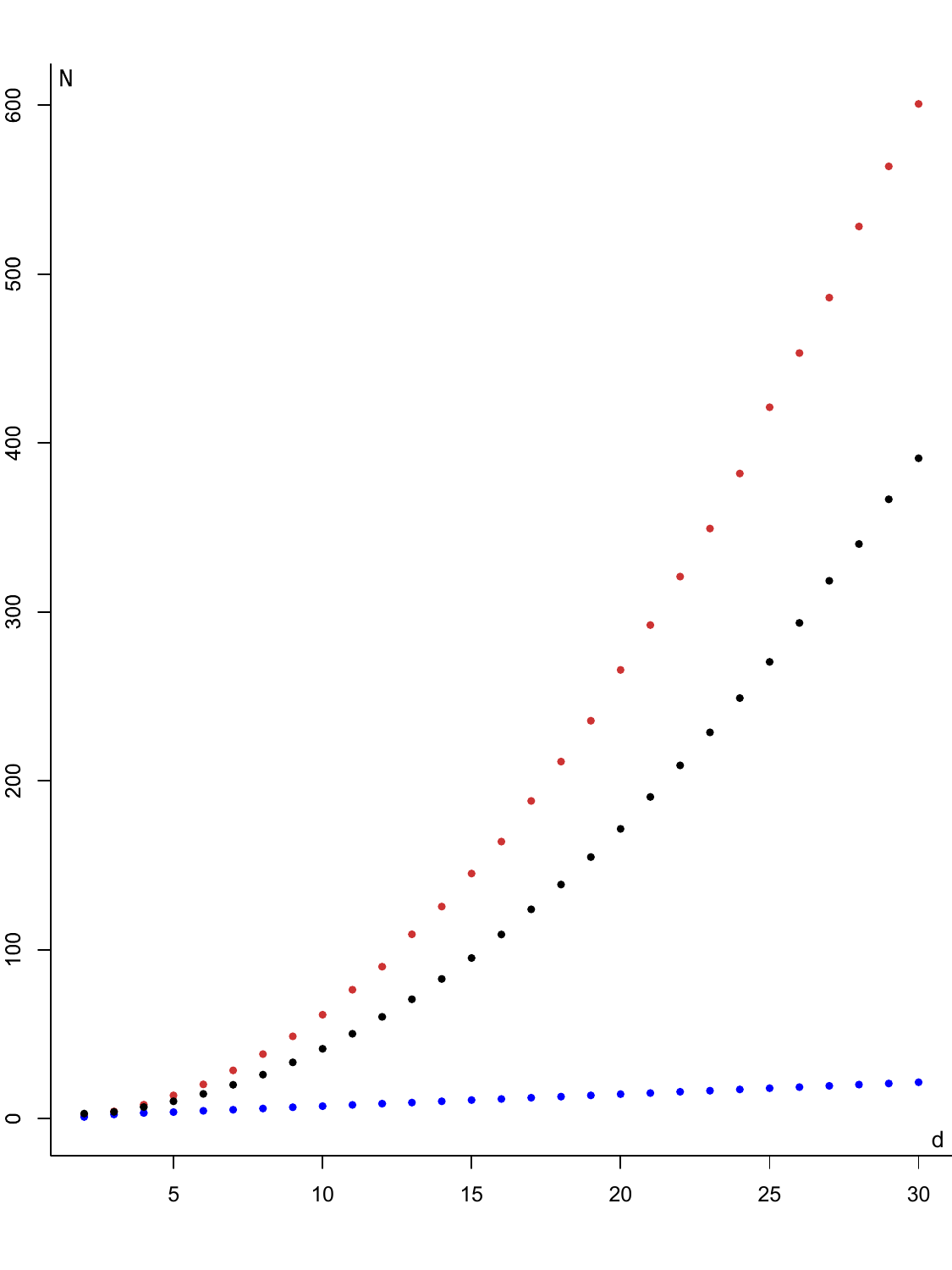}		
		\captionsetup{width=0.95\linewidth}	
		
		\vspace{-4mm}
		
		\caption{\textit{Black dots}: the mean number of steps $\braket{\tilde{N}}$ executed by the algorithm run with parameters $M = 100$, $s = 1$, $\beta = 0.5$. This relationship seems to be quadratic. \textit{Red dots}: the mean number of steps until $\,d_N(U)  =  d$ averaged over the Haar measure. 
			\textit{Blue dots}: the mean number of steps until  $\,d_N(U)  \geq  d/2$ averaged over the Haar measure. These relationships seem to be quadratic and linear, respectively.}

		\label{fig_steps} 
		
	\end{figure}

	We ran this algorithm 1,000 times for $d= 2, \ldots, 30$ with parameters $M = 100$, $\beta=0.5$, $s=1$ and  unitary matrices generated from the Haar distribution. The accuracy was $100\%$ and the average number of executed steps is plotted in Fig. \ref{fig_steps}.

\end{document}